 \pdfoutput=1
\documentclass[%
preprint,
 amsmath,amssymb,
 aps,
floatfix,
]{revtex4-1}
\pdfoutput=1
\usepackage{appendix}
\usepackage{amsmath}
\usepackage{graphicx}
\usepackage{epstopdf}
\usepackage{graphics}
\usepackage{epsfig}
\usepackage[byname]{smartref}
\usepackage{hyperref} 
\usepackage{txfonts}
\usepackage{morefloats}
\usepackage{tocloft}
\usepackage{graphicx}
\usepackage{dcolumn}
\usepackage{bm}
\usepackage{hyperref}

\begin{document}

\preprint{V1}

\title{The next nearest neighbor effect on the 2D materials properties}

\author{Maher Ahmed}
\affiliation{Department of Physics and Astronomy, University of Western Ontario, London ON N6A 3K7, Canada}
\affiliation{Physics Department, Faculty of Science, Ain Shams University, Abbsai, Cairo, Egypt}
\email{mahmed62@uwo.ca}
%



\begin{abstract}
In this work, the effect of introducing next nearest neighbor (NNN) hopping to the 2D materials was
studied using the graphene 2D honeycomb two sublattice as an example. It is
found that introducing NNN to the 2D materials removes the symmetry around
the Fermi level and shifts it, at a small value of NNN hopping. This effect
increases with increasing NNN hopping. If the NNN hopping becomes competitive
with nearest neighbor (NN) hopping, the dispersion relations of the 2D materials changes
completely from NN hopping dispersion relations. The results show that the 2D
material sensitivity for NNN hopping effect is much larger in the 2D
honeycomb lattice than 2D square lattice. This is due to the fact that the
number of NNN sites is equal to six, which is the double of NN sites in the
2D honeycomb lattice. Meanwhile, the number of NNN sites is equal to four
which is equal to NN sites in 2D square lattice. We therefore conclude that
by changing the ratio between NNN and NN sites in the 2D lattice one can tune
the sensitivity for NNN hopping effects.
\end{abstract}

\pacs{Valid PACS appear here}
\maketitle

\def\baselinestretch{1.66}

\section{Introduction}
It is shown in \cite{Selim2011,Ahmed2,Ahmed2,Ahmed3,Ahmed4} that the physical properties of 2D materials like
graphene and magnetic stripes are mainly attributed to both their lattice
structure and range of interactions between its sites. With fixing the range
of interactions to include only the nearest neighbor hopping, a comparison
between the obtained results of the 2D square lattice, zigzag edged, and
armchair edged 2D honeycomb lattice show that the 2D lattice structure and
its edge configuration play very important rule in its dispersion relations
and consequently its possible applications.

Despite this importance of lattice structure in the properties of 2D
materials, experiments and theories
\cite{Barrios-VargasMar2011,Neto1,Deacon2007,Reich2002} show that increasing
the range of the interaction to include the next nearest neighbor (NNN) in
the graphene 2D honeycomb lattice changing its dispersion relations by
removing dispersion symmetry around the Fermi level with shifting Fermi level
value, and changing their behavior around the impurities in the lattice
\cite{PhysRevB.73.241402}. Also, including the next nearest neighbor hopping
in 2D square lattice changing its dispersion relations
\cite{PhysRevLett.106.236803}.

It is interesting to study the effect of introducing the next nearest
neighbor in the structure of the hopping $\mathbf{E}$ matrix, and consequently on the
obtained dispersion relations and the localized edge states of the 2D
materials which has not been previously examined.

In this work, the graphene Hamiltonian \cite{Neto1} which
includes the next nearest neighbor interaction term will be used to study its
effect on the dispersion relations, edge states, and impurities states in the
graphene nanoribbons with zigzag and armchair edge. The obtained results
should also be applicable to the magnetic case, since the next nearest
neighbor interaction term can be added to Heisenberg Hamiltonian
\cite{Xue1996,PhysRevB.38.11898}.

\section{Theoretical model}
The system initially under study is a 2D graphene nanoribbon in the
$xy$-plane. The crystallographic description of graphene honeycomb lattice is
given in \cite{Neto1}. The
nanoribbon is of finite width in the $y$ direction with $N$ atomic rows
(labeled as $n = 1,\cdots,N$) and it is infinite in $x$ direction ($-\infty
\Leftrightarrow \infty$).
\begin{figure}[h]
\centering
\includegraphics[scale=0.2]{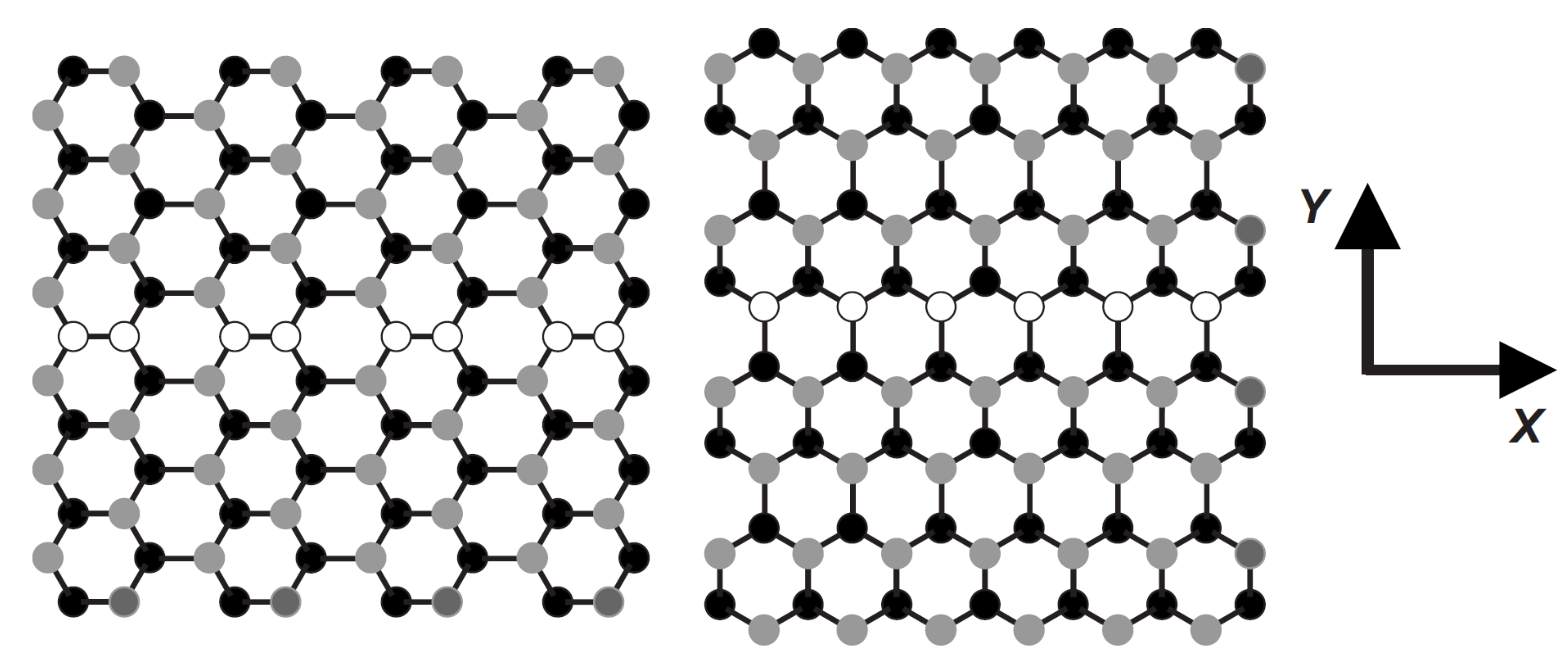}
\caption{Armchair (left) and zigzag (right)  graphene 2D honeycomb nanoribbons in $xy$-plane,  where black (gray) dots are the sublattice
A(B) with a line of impurities (white dots) in the middle of
the sheet, with average spin alignment in $z$ direction. The stripes are finite in $y$ direction with $N$ rows
($n = 1,\cdots,N$) and they are infinite in the $x$ direction.  Figure taken from \cite{rim1}.}\label{fig:graphenelattice6}
\end{figure}
The total Hamiltonian of the system is given by
\begin{equation}\label{secondhamiltonian13}
\hat{H}=-\sum_{\langle ij\rangle} t_{0_{ij}} (\textbf{a}^\dag_i \textbf{b}_j+\textbf{h.c})+t_{1_{ij}}(\textbf{a}^\dag_i \textbf{a}_j+\textbf{b}^\dag_i \textbf{b}_j+\textbf{h.c}),
\end{equation}
where the first term
$t_{0_{ij}}(\approx 2.8 \textrm{eV})$ is the nearest-neighbor hopping energy
and in graphene it is the hopping
between different sublattices $A$ and $B$. Also $t_{1_{ij}} (\approx 0.1
\textrm{eV})$ is the next nearest-neighbor hopping energy which here in
graphene is the hopping in the same sublattice
\cite{Neto1,Deacon2007,Reich2002}. The summations over $i$ and $j$ run over
all the sites where $i$ and $j$ are belong to different sublattice A(B) for
the nearest neighbors hopping term, and they are belonging to the same
sublattice for the next nearest-neighbor hopping energy. Where the nearest neighbor hopping $t_{0_{ij}}$ has a
constant ``bulk" value $t$ when either $i$ and $j$ are in the interior of the
nanoribbon, and another constant value $t_{e}$  when $i$ and $j$ are both at
the edge of the nanoribbon (i.e., in row $n=1$ or $n=N$). Similarly, for the
next nearest-neighbor hopping energy $t_{1_{ij}}$, we assume that it has a
constant value $t'$ when the site $i$ is inside the nanoribbon, and it is
equal to $t'_{e}$ for sites at the edge of the nanoribbon.

Since the nanoribbon extends to $\pm\infty$ in the $x$ direction, we may
introduce a 1D Fourier transform to wavevector $q_x$ along the $x$ direction
for the fermions operators $a^{\dag}_{i}$ ($a_{i}$) and $b^{\dag}_{j}$
($b_{j}$) as follows:
\begin{align}
b_{j}(x)&=\frac{1}{\sqrt{N_0}} \sum_{n} b_{n}(q_x) e^{-i \mathbf{q}_x\cdot \mathbf{r}_j} \hspace{20pt}
b_{j}^\dag(x)=\frac{1}{\sqrt{N_0}} \sum_{n} b^\dag_{n}(q_x) e^{i \mathbf{q}_x\cdot \mathbf{r}_j}\label{far33}\\
a_{i}(x)&=\frac{1}{\sqrt{N_0}} \sum_{n} a_{n}(q_x) e^{-i \mathbf{q}_x\cdot \mathbf{r}_i} \hspace{20pt}
a_{i}^\dag(x)=\frac{1}{\sqrt{N_0}} \sum_{n} a^\dag_{n}(q_x) e^{i \mathbf{q}_x\cdot \mathbf{r}_i}. \nonumber
\end{align}
Here $N_0$ is the (macroscopically large) number of spin sites in any row,
$\mathbf{q_x}$ is a wavevector in the first Brillouin zone of the reciprocal
lattice and both $\mathbf{r}_i$ and $\mathbf{r}_j$ is the position vectors of
any carbon sites $i$ and $j$. The new operators obey the following
commutation relations:
\begin{equation}\label{comu33}
\left[a_{n}(q_x),a^\dag_{n}(q'_x)\right]=\delta_{q_xq'_x},\hspace{30pt}\left[b_{n}(q_x),b^\dag_{n}(q'_x)\right]=\delta_{q_xq'_x}.
\end{equation}

Also, we define the hopping sum:
\begin{eqnarray}
\nonumber
 \tau(q_x) &=&\sum_{\nu} t_{{0}_{ij}}  e^{-i\mathbf{q}_x \cdot (\mathbf{r}_i-\mathbf{r}_j)}.\\
 \tau'(q_x) &=& \sum_{\nu'} t_{{1}_{ij}}  e^{-i\mathbf{q}_x \cdot (\mathbf{r}_i-\mathbf{r}_j)} \label{hoppingsum}.
\end{eqnarray}
The sum for the hopping terms $t_{{0/1}_{ij}}$ is taken to be over all $\nu$
nearest neighbors and over all $\nu'$ next nearest-neighbor in the lattice
which depends on the edge configuration as zigzag or armchair for the stripe. For the armchair
configuration, the hopping sum for nearest neighbors gives the following
factors $ \tau_{nn'}(q_x)$
\begin{eqnarray}
  \tau_{nn'}(q_x)=t\left[\exp(iq_xa)\delta_{n',n}+\exp\left(i\frac{1}{2}q_xa\right)\delta_{n',n\pm1}\right]
\end{eqnarray}
and for the zigzag configuration, it gives
\begin{eqnarray}
  \tau_{nn'}(q_x)=t\left[2\cos\left(\frac{\sqrt{3}}{2}q_xa\right)\delta_{n',n\pm1}+\delta_{n',n\mp1}\right].
\end{eqnarray}

The hopping sum for next nearest neighbors gives the following factors
$\tau'_{nn'}(q_x)$
\begin{eqnarray}
  \tau_{nn'}(q_x)=t'\left[\delta_{n',n\pm2}+ 2\cos(q_x a3/2)\delta_{n',n\pm1}\right]
\end{eqnarray}
for the armchair configuration, and
\begin{eqnarray}
  \tau_{nn'}(q_x)=2t'\left[\cos(\sqrt{3}q_x a)\delta_{n',n}+\cos(\sqrt{3}q_x a/2)\delta_{n',n\pm2}\right]
\end{eqnarray}
for the zigzag configuration case, where the $\pm$ sign, in all the above
factors, depends on the sublattice since the atom line alternates from A and
B.

Substituting Equations~\eqref{far33} and \eqref{hoppingsum}  in
Equation~\eqref{secondhamiltonian13}, and rewriting the summation over
nearest and next nearest neighbors sites, we get the following form of the
operator term $\hat{H}$:
\begin{eqnarray}
   \hat{H}&=&-\sum_{nn'}  \tau'(q_x) \left(a^\dag_{n} a_{n'}+ b^\dag_{n} b_{n'} \right)   + \tau(q_x) a_{n}  b^\dag_{n'}+ \tau(-q_x) a^\dag_{n}  b_{n'}.
\end{eqnarray}
The first terms count the elementary excitations on each sublattice, while
the second describes the coupling between the sublattices.

In order to diagonalize $\hat{H}$ and obtain the dispersion relations for
graphene nanoribbons, we may consider the time evolution of the creation and
the annihilation operators $a^{\dag}_{i}$ ($a_{i}$) and $b^{\dag}_{j}$
($b_{j}$), as calculated in the Heisenberg picture in quantum mechanics where
the time dependent is transferred from the system wavefunction to the
operators. In this case, the equations of motion (using the units with
$\hbar=1$) for the annihilation operators $a_{i}$($b_j$) are as follows
\cite{Bes2007,Liboff,Kantorovich2004,Roessler2009,HenrikBruus2004}:
\begin{eqnarray}
\frac{d a_{n}}{dt} &=&i[H,a_{n}]  \nonumber\\
    &=&i\sum_{nn'}-\tau'(q_x)  a_{n'}-\tau(-q_x) b_{n'} \label{ch3em13}
\end{eqnarray}
and
\begin{eqnarray}
 \frac{d b_{n}}{dt} &=&i[H,b_{n}]  \nonumber\\
      &=&i\sum_{nn'}-\tau'(q_x)  b_{n'}-\tau(q_x) a_{n'} \label{ch3em23}
\end{eqnarray}
where the commutation relation in Equation~\eqref{comu33} was used, as well
as the operator identity $[AB,C]=A[B,C]+[A,C]B$.

The electronic dispersion relations of the graphene  (i.e., energy or
frequency versus wavevector) can now be obtained by solving the above
operator equations of motion. The electronic energy can be expressed in terms
of the frequency using the relation $E = \hbar\omega$, and assuming that
electronic energy modes behave like $\exp[-i\omega(q_x)t]$, on substituting
this time dependent in Equations \label{ch3em13} and \label{ch3em23}, we get
the following sets of coupled equations:
\begin{eqnarray}
\omega(q_x) a_{n}  &=&\sum_{nn'}\tau'_{nn'}(q_x)  a_{n'}+\tau_{nn'}(-q_x) b_{n'}  \\
& &  \nonumber\\
\omega(q_x) b_{n}  &=&\sum_{nn'}\tau_{nn'}(q_x) a_{n'}+\tau'_{nn'}(q_x)  b_{n'}
\end{eqnarray}
The above equations can be written in matrix form as following
\begin{eqnarray}
 \omega(q_x)\left[%
\begin{array}{c}
a_{n}   \\
b_{n}   \\
\end{array}%
\right]  &=& \left[%
\begin{array}{cc}
  T'(q_x)   & T(q_x)  \\
  T^*(q_x) & T'(q_x)    \\
\end{array}%
\right] \left[%
\begin{array}{c}
a_{n}   \\
b_{n}   \\
\end{array}%
\right]\label{eq3eqing6}
\end{eqnarray}

where the solution of this matrix equation is given by the condition
\begin{eqnarray}
\det \left[%
\begin{array}{cc}
 -(\omega(q_x)I_N -T'(q_x))   & T(q_x)  \\
  T^{*}(q_x) & -(\omega(q_x)I_N- T'(q_x))   \\
\end{array}%
\right]=0 \label{matrix23}
\end{eqnarray}

\begin{table}[h]
 \caption{Nearest neighbor hopping matrix elements for the graphene as 2D honeycomb
lattice}\label{311}
\begin{tabular}{lcl}
  \hline\hline
  Parameter& Zigzag               & Armchair \\ \hline
   $\alpha$&         0           &$te^{-iq_xa}$  \\
   $\beta$ &$2t \cos(\sqrt{3}q_x a/2)$&$te^{iq_xa/2}$\\
   $\gamma$& $t$ &$ te^{iq_xa/2}$ \\
  \hline
\end{tabular}
  \centering
 \caption{Next nearest neighbor hopping matrix elements for the graphene as 2D honeycomb
lattice}\label{312}
\begin{tabular}{lcll}
  \hline\hline
  Parameter& Zigzag  &  Parameter           & Armchair \\ \hline
  $\epsilon$& $2t' \cos(\sqrt{3}q_x a)$& $\theta$                  & $t'$ \\
  $\zeta$& $2t' \cos(\sqrt{3}q_x a/2)$ &$\eta$&$2t' \cos(q_x a3/2)$ \\
  \hline
\end{tabular}
  \centering
\end{table}

Where $T(q_x)$ and $T'(q_x)$ are nearest and next nearest exchange matrices
respectably, which are depend on the orientation of the ribbon, and
$\omega(q_x)$ are the energies of the modes. The matrix $T(q_x)$ is given by
\begin{equation}\label{}
\left(
  \begin{array}{ccccc}
    \alpha & \beta   &     0 & 0 & \cdots \\
    \beta   & \alpha & \gamma& 0 & \cdots \\
    0       & \gamma  &\alpha&  \beta    &\cdots \\
    0       &      0  & \beta & \alpha & \cdots \\
     \vdots & \vdots& \vdots & \vdots & \ddots \\
  \end{array}
\right).
\end{equation}

The matrix $T'(q_x)$ for zigzag ribbon is given	by
\begin{equation}\label{}
\left(
  \begin{array}{cccccc}
    \epsilon& 0     &\zeta & 0    & 0 & \cdots \\
   0   &\epsilon & 0   & \zeta    & 0 & \cdots \\
    \zeta     &0 &\epsilon &  0 & \zeta &\cdots \\
    0     &  \zeta  & 0 & \epsilon& 0 & \cdots \\
    0     &   0 &   \zeta  & 0 & \epsilon & \cdots \\
     \vdots & \vdots& \vdots & \vdots & \vdots& \ddots \\
  \end{array}
\right)
\end{equation}
and the matrix $T'(q_x)$ for armchair ribbon is given by
\begin{equation}\label{}
\left(
  \begin{array}{cccccc}
    0& \eta    &\theta & 0    &0  & \cdots \\
   \eta   &0 & \eta  & \theta   &  0& \cdots \\
    \theta   &\eta &0 &  \eta & \theta &\cdots \\
    0     &  \theta& \eta & 0& \eta & \cdots \\
    0     &   0 &   \theta  & \eta& 0 & \cdots \\
     \vdots & \vdots& \vdots & \vdots & \vdots& \ddots \\
  \end{array}
\right)
\end{equation}
The parameters $\alpha,\beta,\gamma,\epsilon,\zeta,\theta$ and $\eta$ depend
on the stripe edge geometry and are given in Tables \ref{311} and \ref{312}.

\subsection{Neglecting the next nearest neighbor hopping as special case}
The next nearest neighbor hopping $t'$ can be neglected compared to nearest
neighbor hopping $t$, in this case the $T'(q_x)$ is equal to zero matrix
$\mathbf{0}$ and Equation~\eqref{eq3eqing6} become as following
\begin{eqnarray}
\det \left[%
\begin{array}{cc}
 -(\omega(q_x)I_N )   & T(q_x)  \\
  T^{*}(q_x) & -(\omega(q_x)I_N)   \\
\end{array}%
\right]=0 \label{matrix23}
\end{eqnarray}
which is the result obtained before for graphene ribbons using the tight
binding model with neglecting NNN hopping \cite{rim1}. It is also very
similar to the case of magnetic stripes in \cite{Ahmed2}, which do not have NNN
exchange, the only difference between the magnetic case and TBM graphene
without NNN hopping is the effect of $\alpha$, i.e. insite energy.

\section{Numerical calculations}
The dispersion relations for the above graphene nanoribbons are obtained
numerically as the eigenvalues \cite{algebra,RefWorks:27} for the matrix
Equation \eqref{eq3eqing6}. This is very similar to graphene with only (NN) \cite{Ahmed2}, and therefore the same numerical calculations method used
there will be used here to get its solutions.

\section{Results}
To compare our results for NN and NNN interaction with the tight-binding
Hamiltonian, with only NN interactions, we choose our stripes sizes, scaling
our result to be dimensional less quantities, and choose physical parameters
matched that ones used  in reference \cite{rim1} for
graphene.
\begin{figure}[hp]
  \centering
  \begin{tabular}{cc}
\includegraphics[scale=.6]{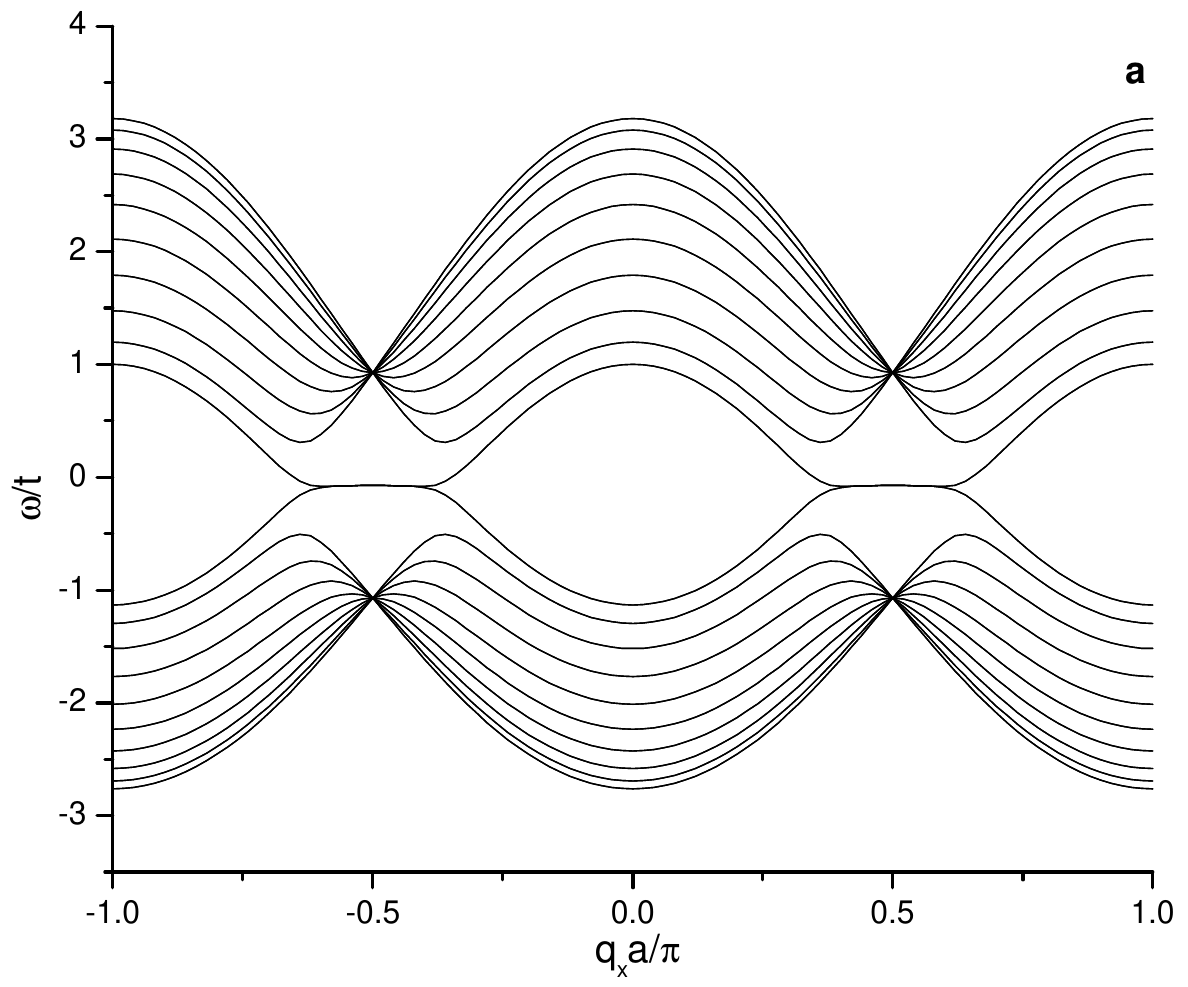}& \includegraphics[scale=.6]{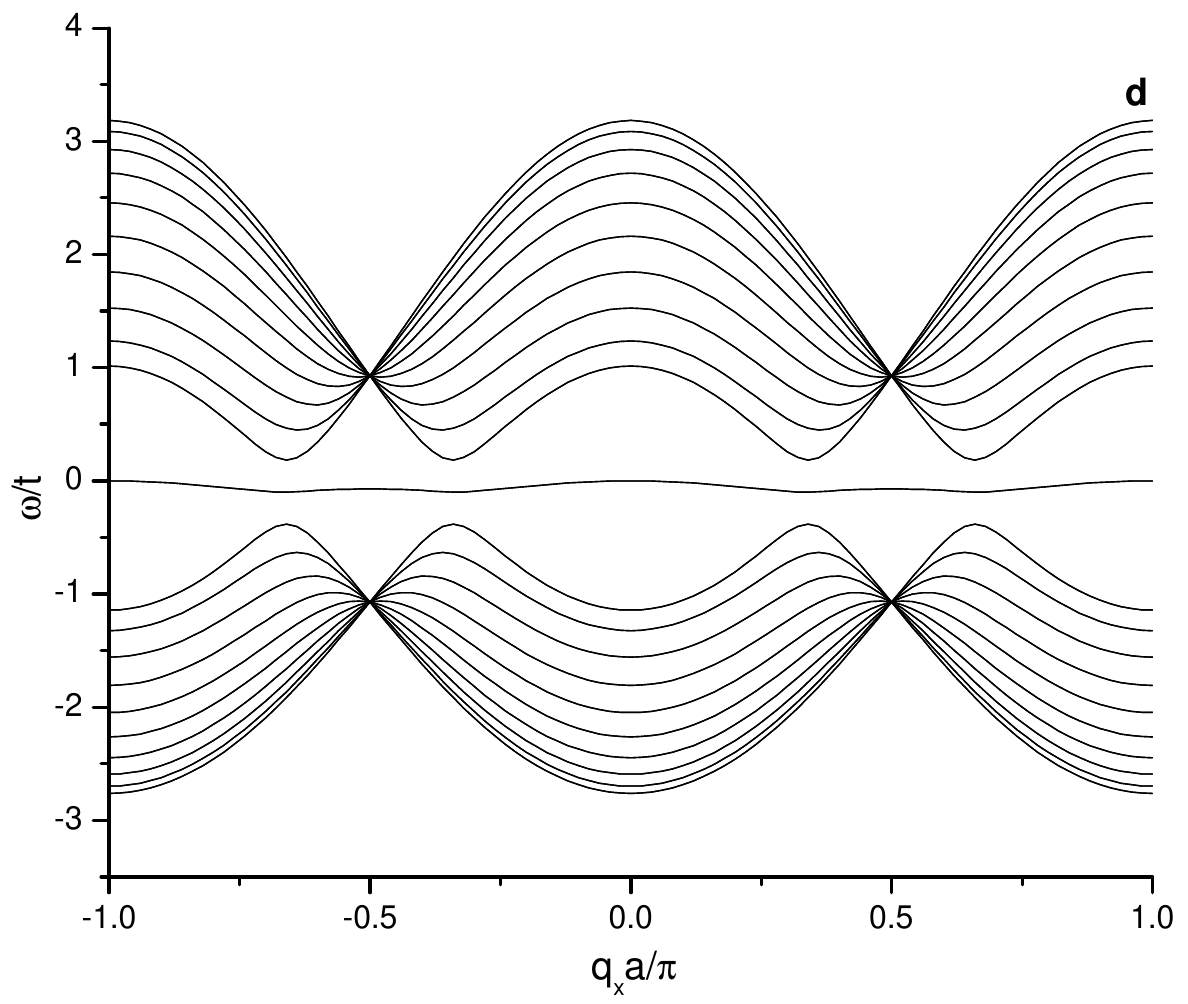}\\
\includegraphics[scale=.6]{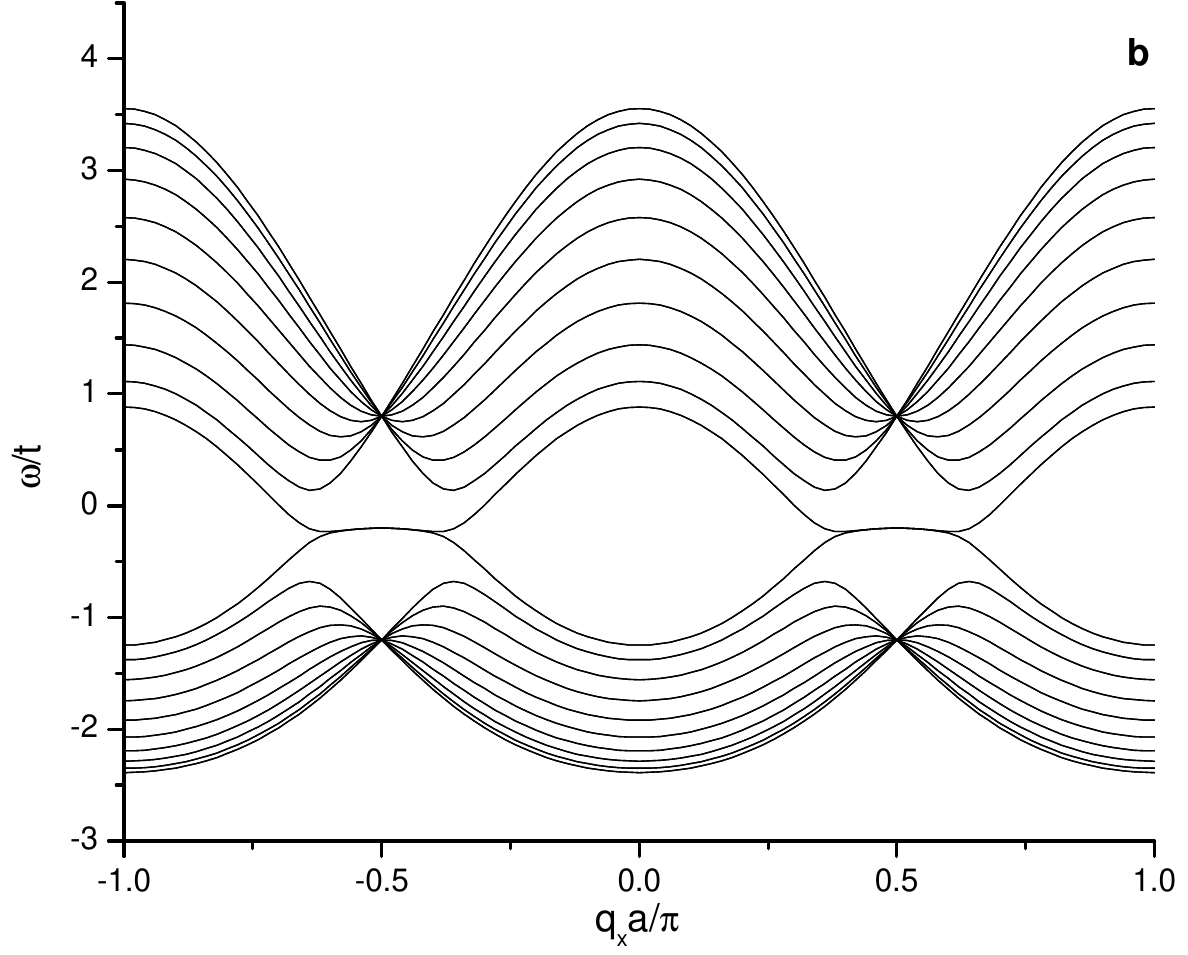}&\includegraphics[scale=.6]{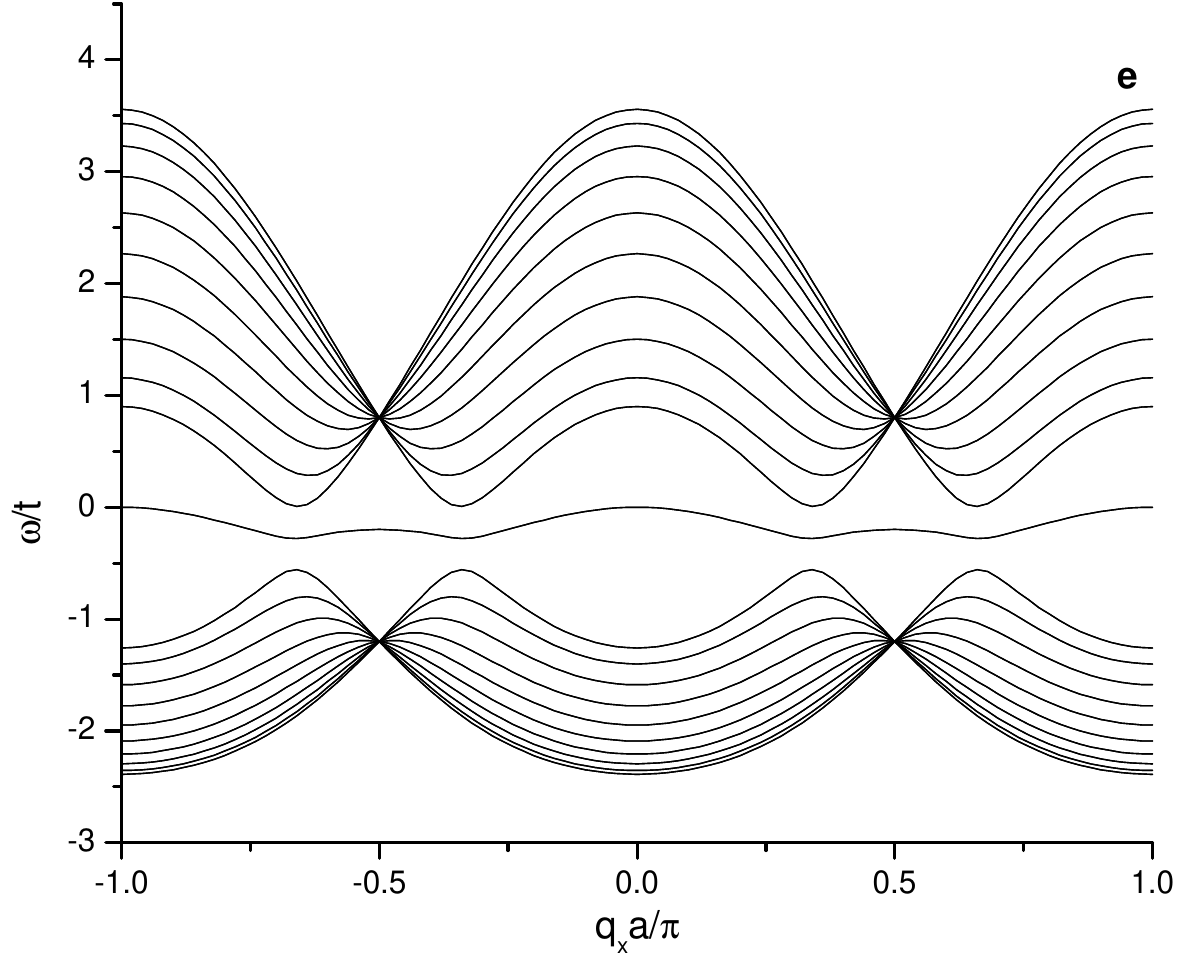}\\
\includegraphics[scale=.6]{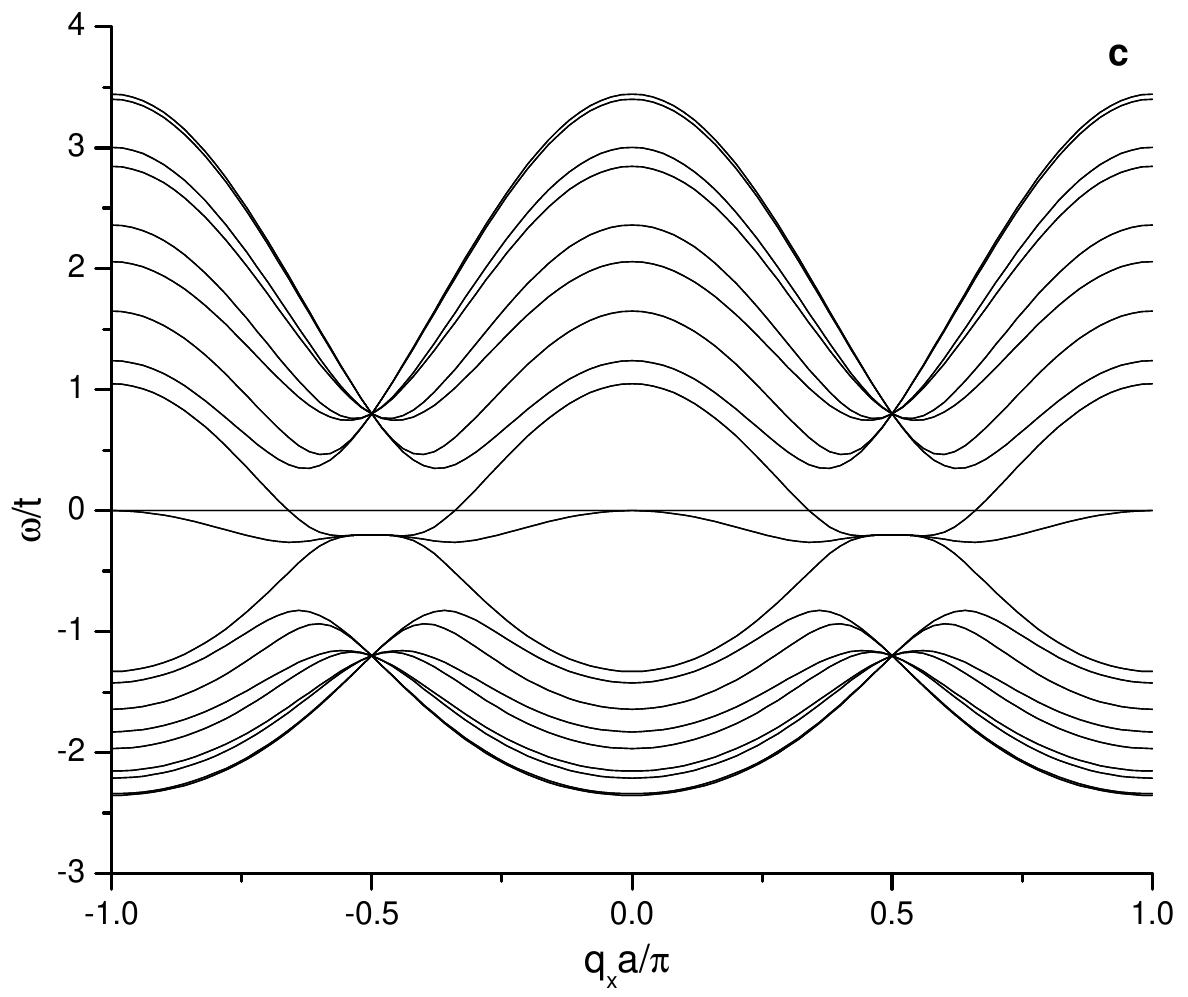}&\includegraphics[scale=.6]{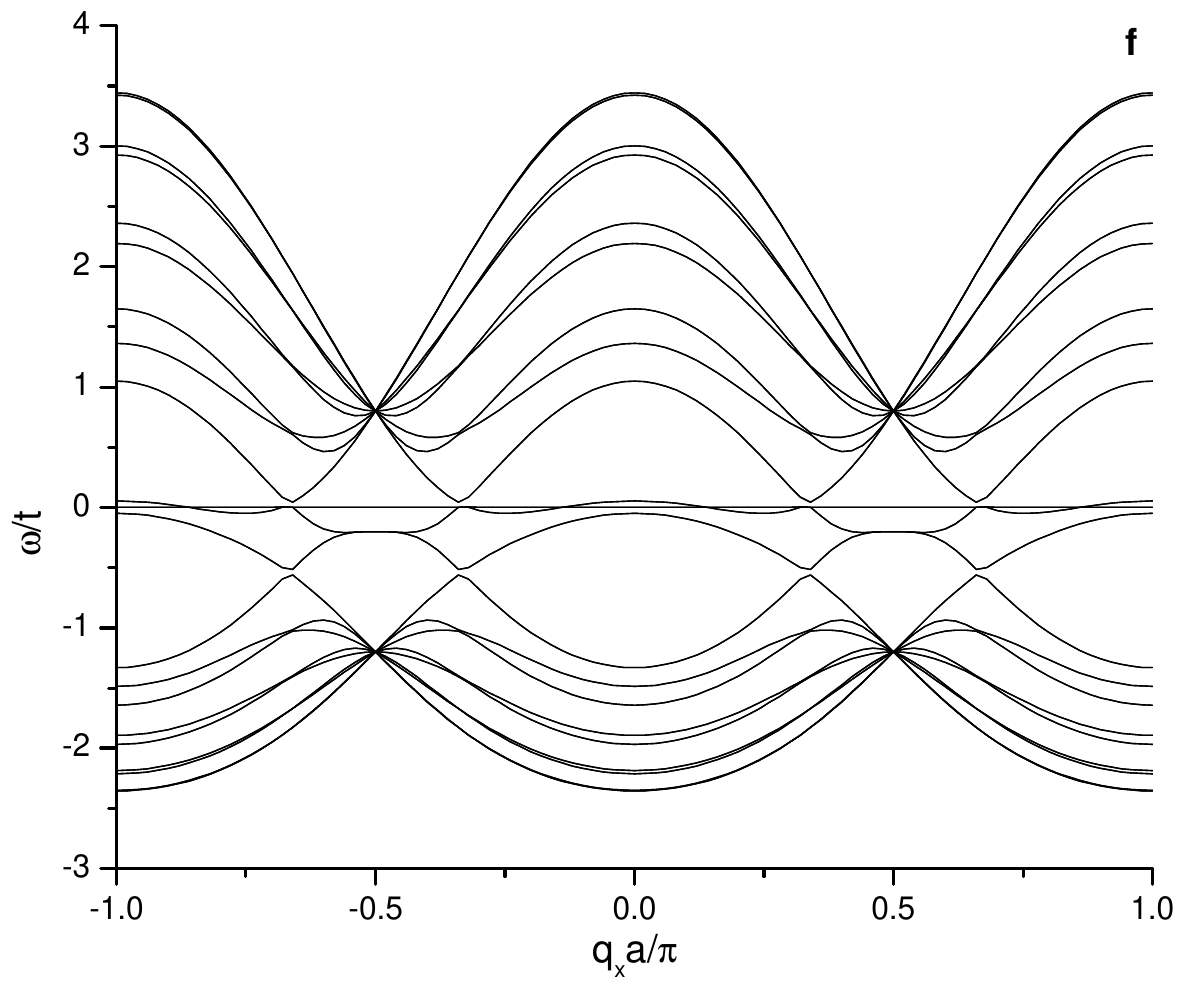}
\end{tabular}
  \caption{The effect of next nearest
neighbor interaction in the dispersion relations, edge
states, and impurities states in the graphene zigzag nanoribbons. Right side stripe width $N=20$ (a) $t'=0.036t$ (b) $t'=0.1t$  (c) $t'=0.1t$ and with impurities line at
row number 11 with $J_I=0$. Left side stripe width $N=21$ (d) $t'=0.036t$ (e) $t'=0.1t$  (f) $t'=0.1t$ and with impurities line at
row number 11 with $J_I=0$.}\label{zigzag6}
\end{figure}

\begin{figure}[hp]
  \centering
  \begin{tabular}{cc}
\includegraphics[scale=.6]{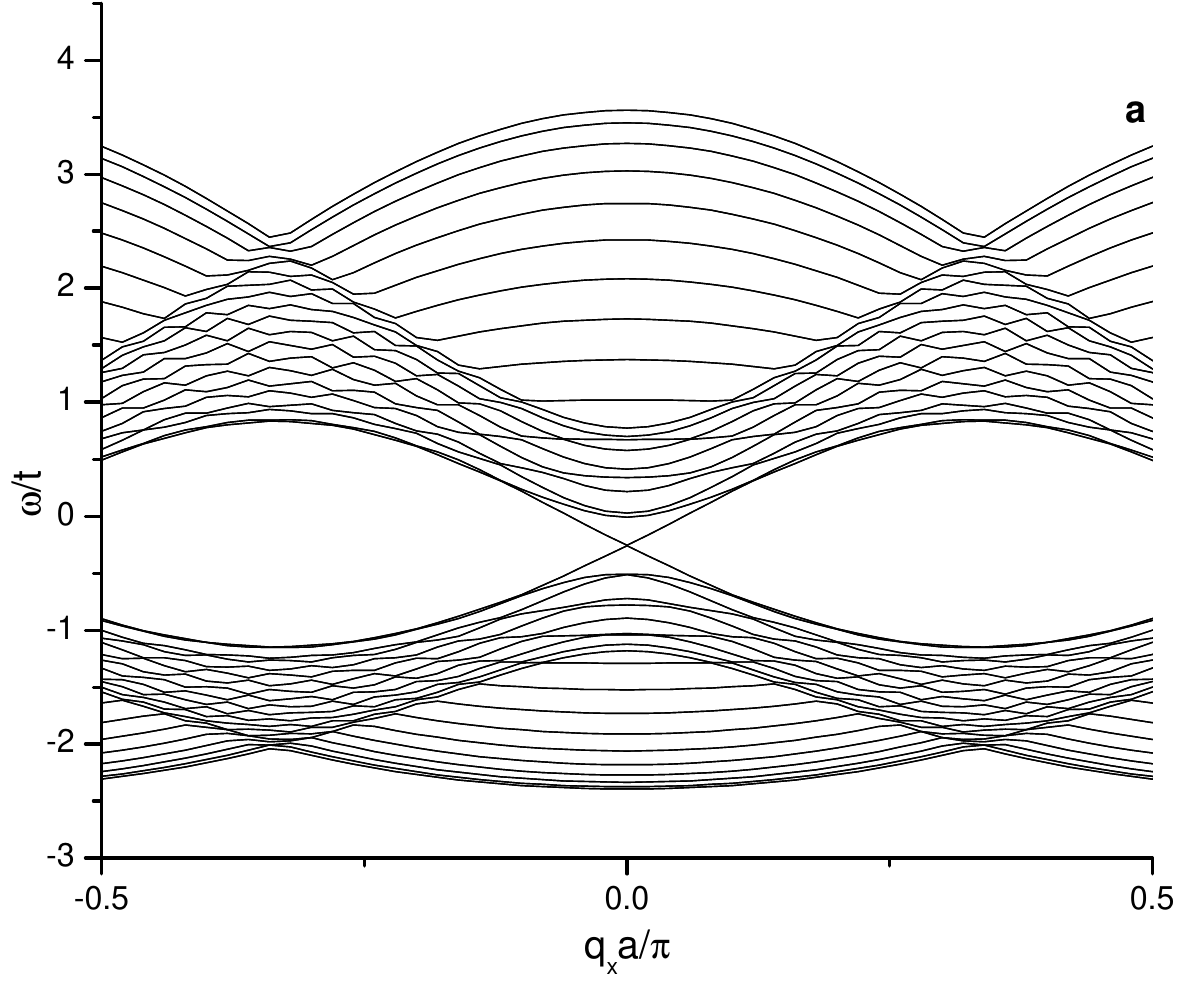}& \includegraphics[scale=.6]{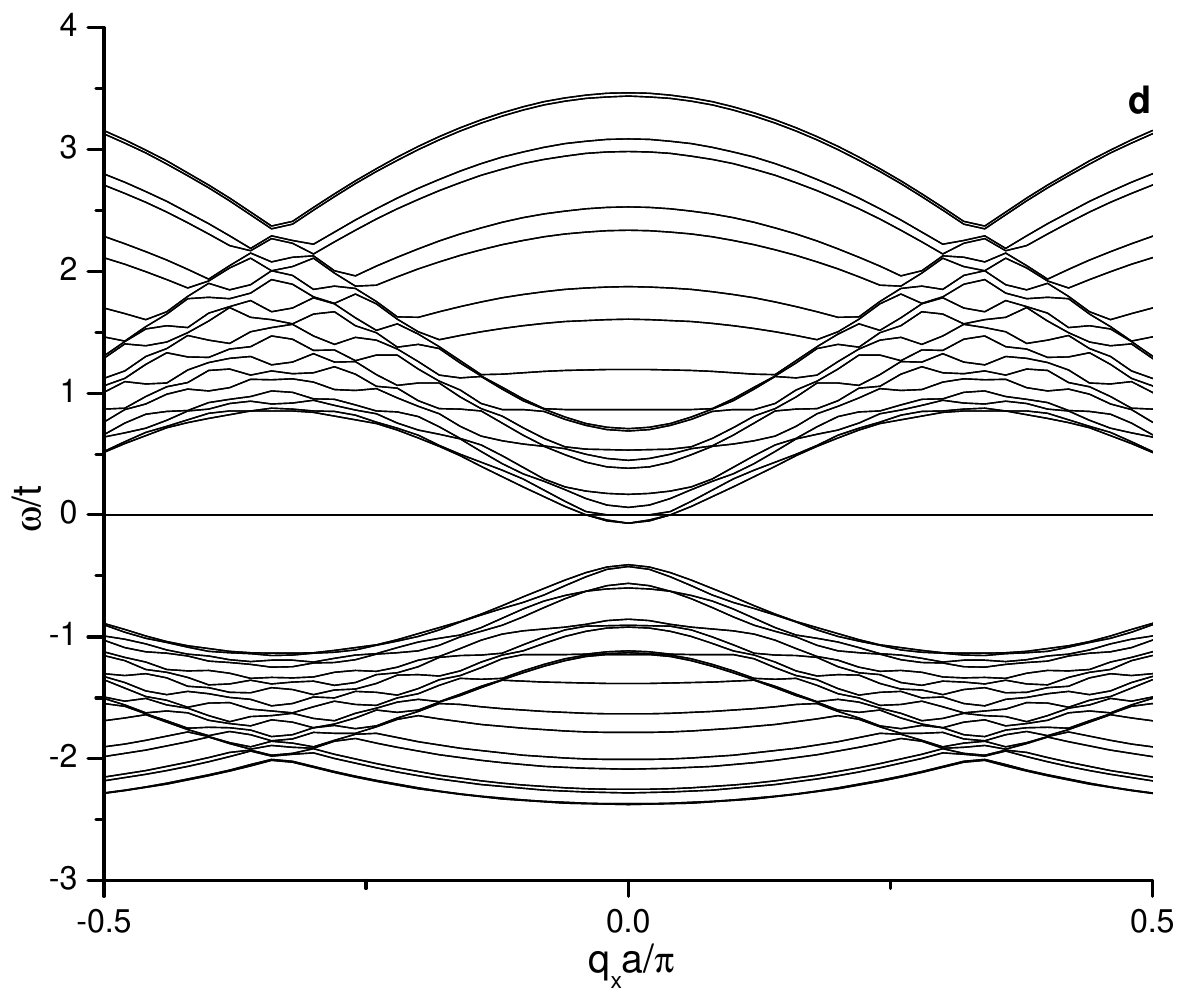}\\
\includegraphics[scale=.6]{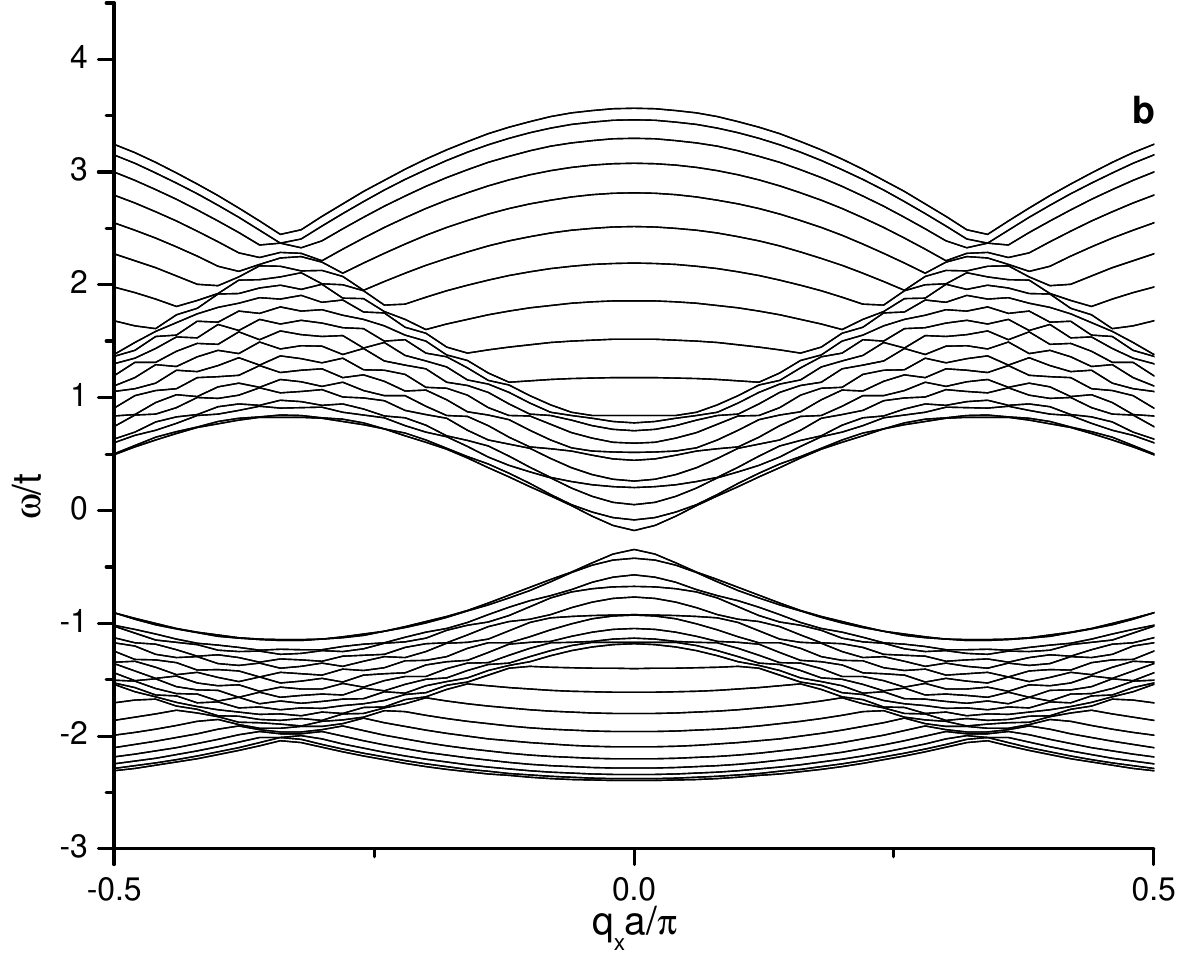}&\includegraphics[scale=.6]{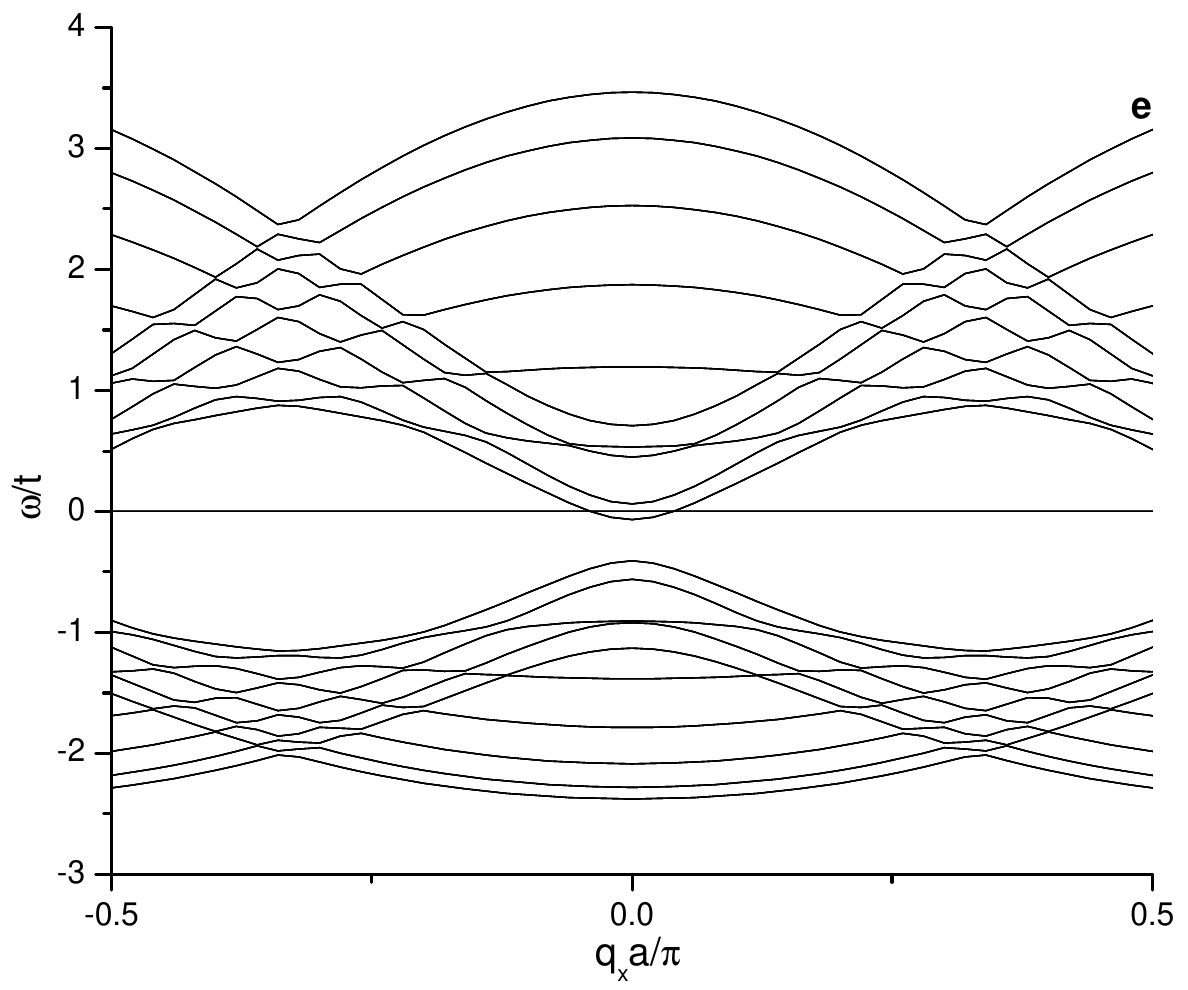}\\
\includegraphics[scale=.6]{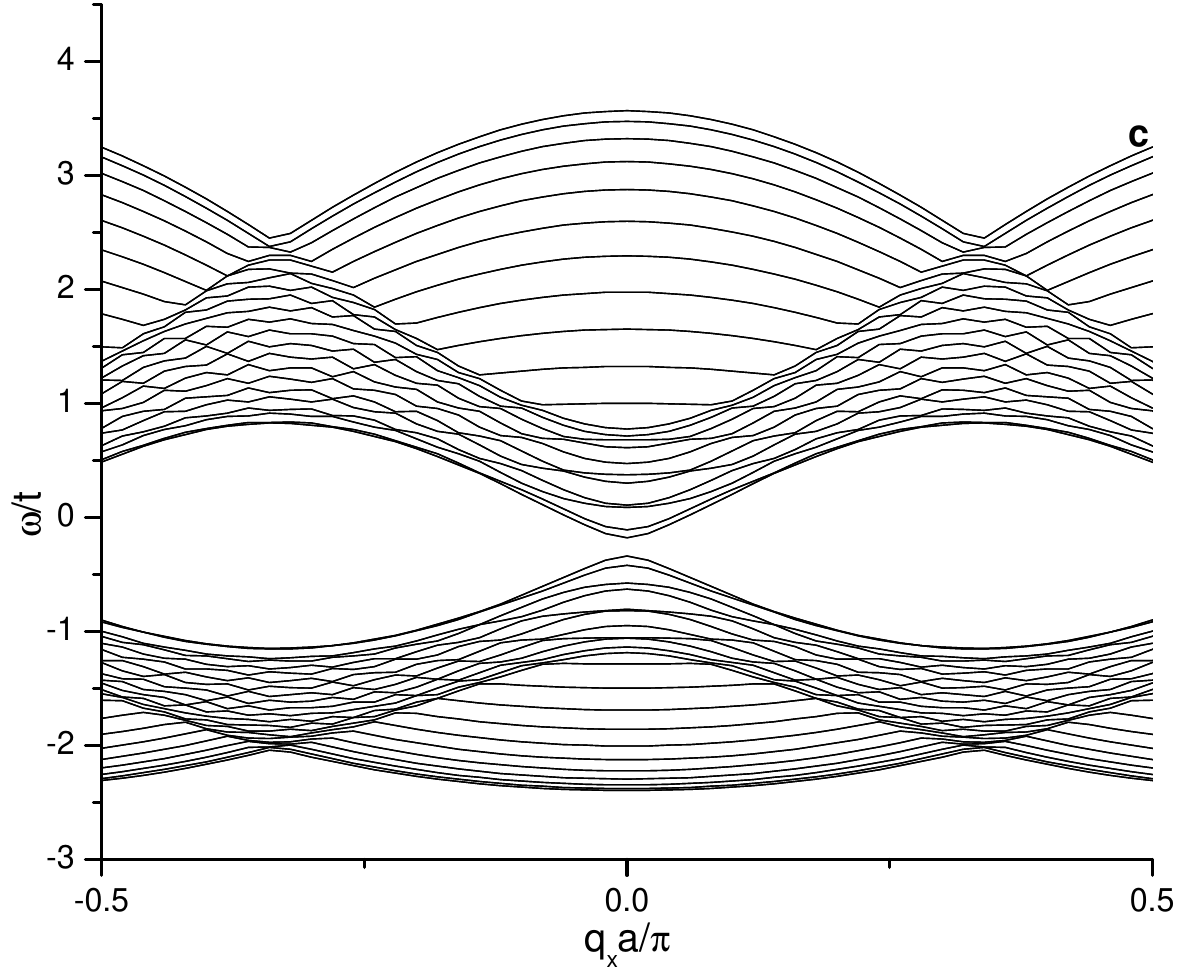}&\includegraphics[scale=.6]{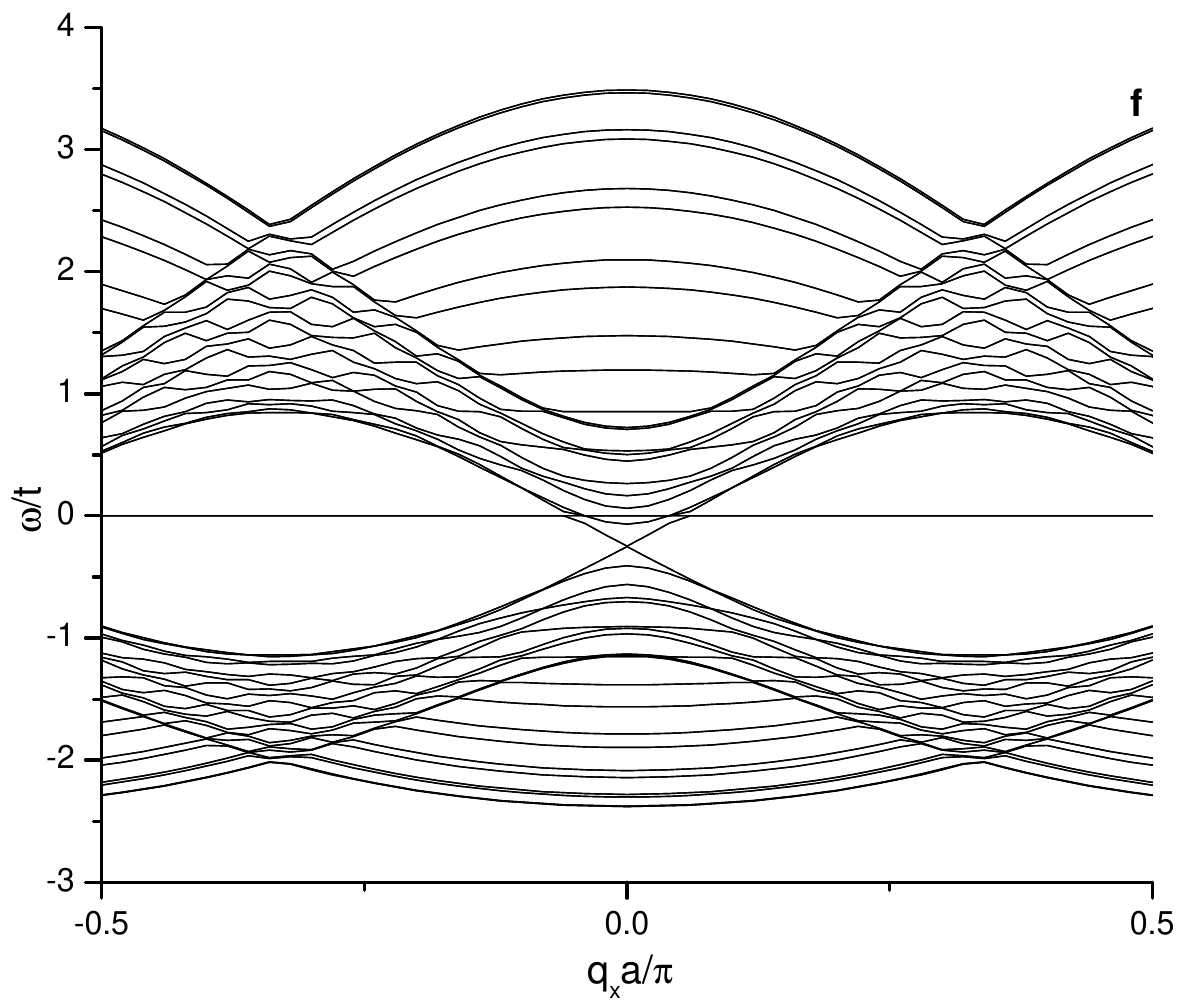}
\end{tabular}
  \caption{The effect of next nearest
neighbor interaction in the dispersion relations, band gap, and impurities states in the graphene armchair nanoribbons. Right side $t'=0.1t$ for stripe width (a) $N=20$ (b) $N=21$  (c) $N=22$. Left side $t'=0.1t$ and with impurities line at
row number 11 with $J_I=0$ for stripe width (a) $N=20$ (b) $N=21$  (c) $N=22$.}\label{armchair6}
\end{figure}

Figures \ref{zigzag6} shows the effect of next nearest neighbor interaction
in the dispersion relations, edge states, and impurities states in the
graphene zigzag nanoribbons, as expected all Figure show the removing
dispersion symmetry around the Fermi level with shifting the Fermi level
value toward valance band.  Figure \ref{zigzag6} (a) and (b) show the effect
for the next nearest neighbor value of $t'=0.036t$, which correspond to
$t'\approx 0.1$ eV and $t\approx2.8$ eV given in references
\cite{Neto1,Deacon2007}, The changing in the dispersion symmetry around the
Fermi level with shifting the Fermi level value toward valance band is small
compared by the obtained results in \cite{Ahmed2} for the same zigzag nanoribbons
without NNN interaction, as the value of NNN increases to $t'=0.1t$ as shown
in Figure \ref{zigzag6} (c) and (d) the density of states increase in the
conduction band (high energy levels) and decreasing in the valance band (low
energy levels) removing the symmetry around the Fermi level and shifting it.
The Figures show that including NNN effecting the flatness of the edge
localized states of zigzag graphene nanoribbons but not effecting its
position in Fermi Level, as the NNN increase the flatness decreases which
reflect the introducing of $q_x$ depends for hopping in edge sites, which
more clear for the extended localized edge state in zigzag with width $N=21$.

Figures \ref{zigzag6} (e) and (f) show the effect of NNN on the impurities
states in the zigzag nanoribbons. It is clear that the position of energy
state of impurities line not affected by including NNN, which is a result of
not participating for the impurities in NNN hopping in this calculation. But
introducing the NNN hopping in the lattice shifting the Fermi level and
changing the density of the states around the impurities level. As, NNN
increase the impurities level move to more density of states region, this
explain the appearance of impurities level as a moving peak in the density of
states for the graphene, with increasing surrounding density of states as NNN
increasing \cite{PhysRevLett.106.236803}.

Figures \ref{armchair6} show the effect of next nearest neighbor interaction
in the dispersion relations, and impurities states in the graphene armchair
nanoribbons. The behavior in the armchair case with including NNN is very
similar to the zigzag case given above for removing the symmetry around the
Fermi level, shifting it, and its effect in the impurities level relative
position to Fermi level. There is no any effect on the absence of edge states
in armchair nanoribbons and on their band shape at $q_xa/\pi=0$ with
including NNN in the model.

\section{Discussion and Conclusions}
In this work, the effect of introducing NNN hopping to the 2D materials was
studied using the graphene 2D honeycomb two sublattice as example. Including
the NNN in the model add NNN hopping matrix $T'(q_x)$, which depending on the
momentum $q_x$ in the direction of nanoribbons symmetry, to the diagonal sub
matrices $\alpha I_N$ in the $\mathbf{E}$ matrix. This shows that NNN hopping
matrix $T'(q_x)$ is real and is describing the hopping with translation
motion in the sublattice sites in both zigzag and armchair stripes.

\begin{figure}[h!]
  \centering
  \begin{tabular}{cc}
\includegraphics[scale=.6]{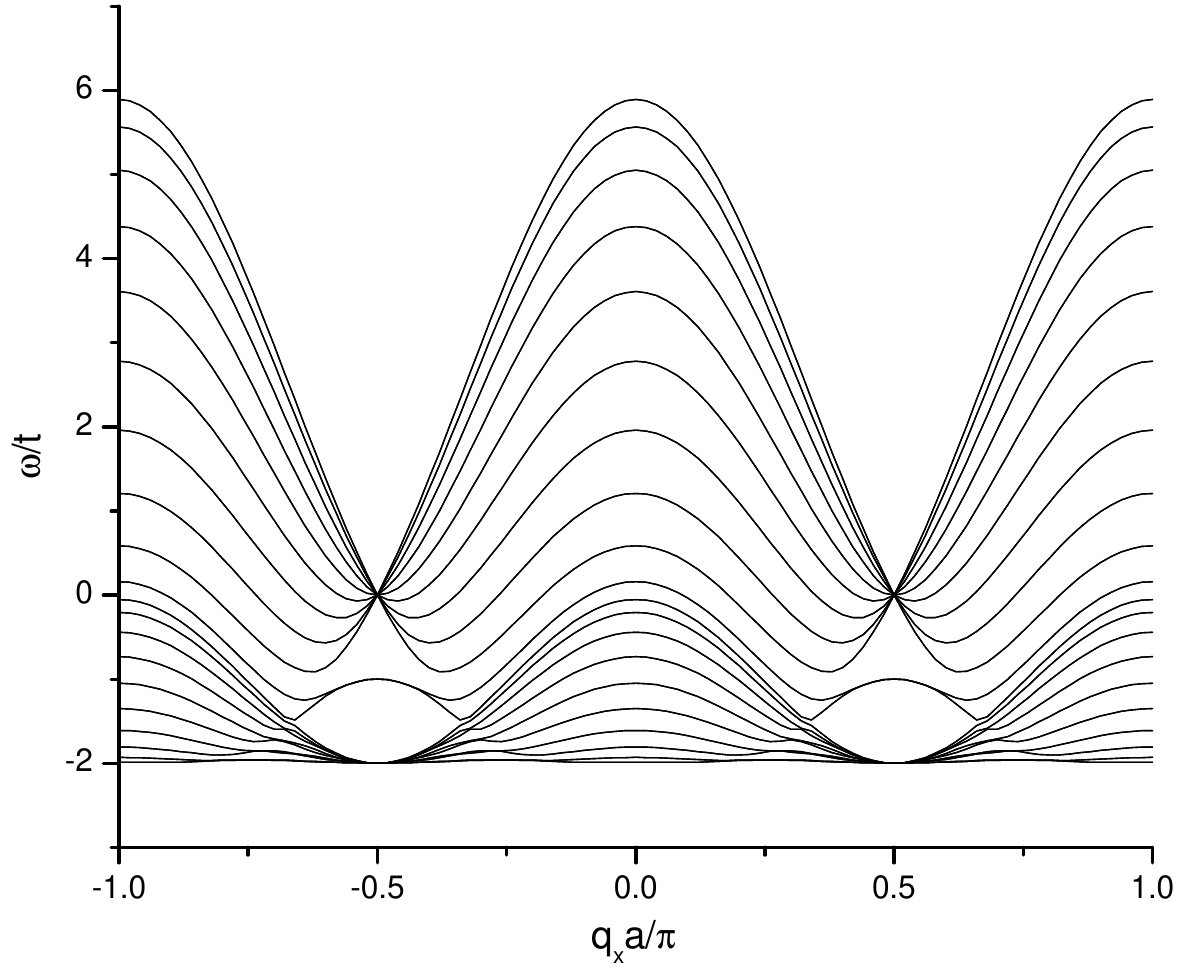}& \includegraphics[scale=.6]{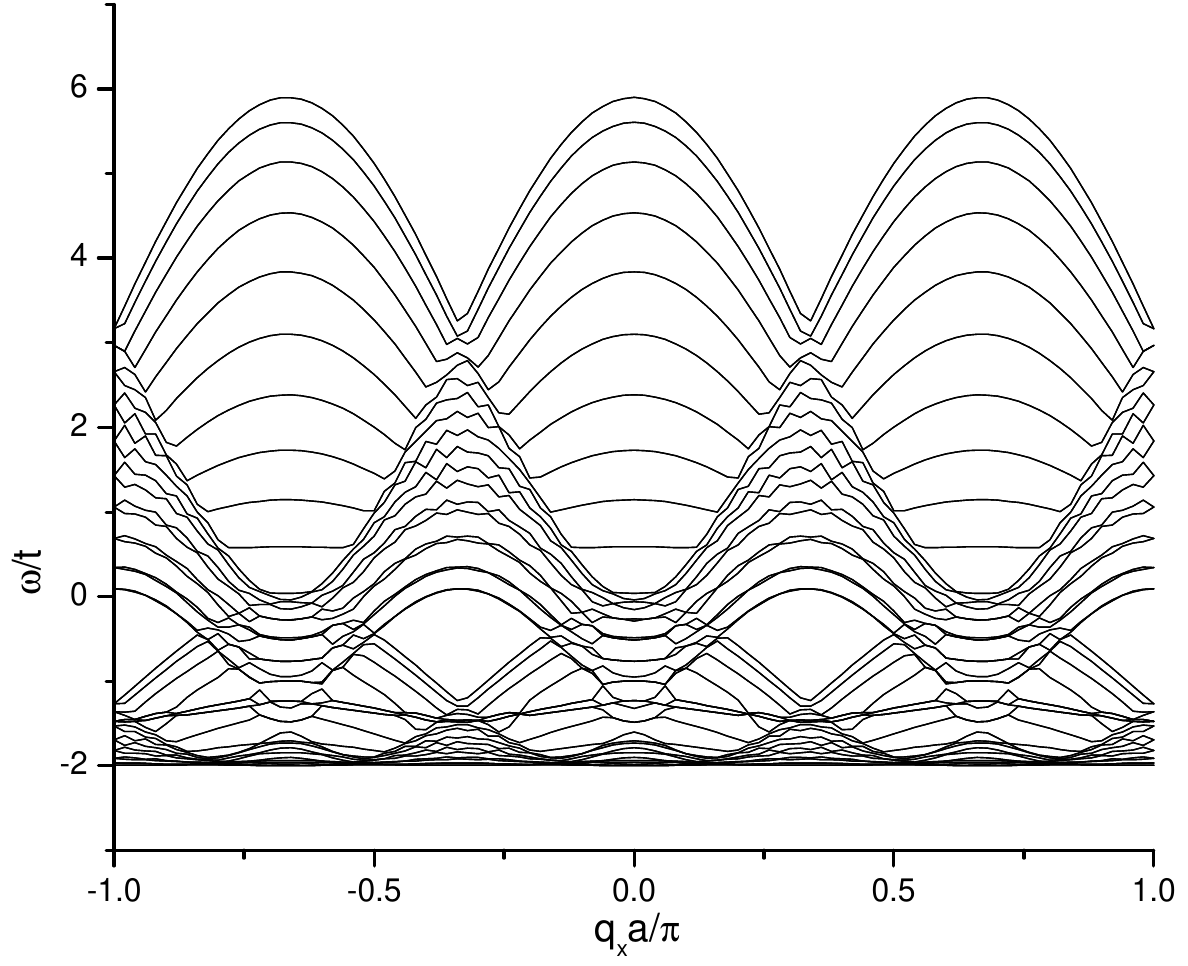}
\end{tabular}
  \caption{The effect of high next nearest
neighbor interaction $t'=0.5t$ in the dispersion relations on zigzag left side and armchair right side lattice with width $N=20$.}\label{square2}
\end{figure}
When $t'$ is equal to very small percentage of $t$, the probability is very
small for the particles to hopping in the same sublattice by NNN hopping and
consequently the net number of particles in NNN hopping is very small which
result a small effect in the dispersion relations of only NN hopping. As the
percentage increases, the probability for NNN hopping increases, and
consequently the net number of particles in NNN hopping increases. This
results in an increasing effect in the dispersion relations of only NN
hopping. The main effect of NNN hopping in small range is changing the
density of states for dominated NN hopping dispersion relations, which can be
explained as following: since the probability for the particles to hopping in
the same sublattice by NNN hopping is increasing with increasing its energy
and consequently the net number of particles from every mode that able to do
NNN hopping is proportional to the mode energy. The highest energy mode has
the highest percentage number of particles that participating in NNN hopping,
this percentage of particles decreases with decreasing the energy of the
mode, most of this NNN particles will be trapped in low energy modes. The
overall effect is the available particles densities is lowest in high energy
modes and highest in low energy modes. This means that the available momentum
spaces in high energy modes is increased for particles in NN hopping due to
NNN effect consequently the density of states is increased in high energy
modes, while the available momentum spaces in low energy modes is decreased
for particles in NN hopping due to NNN effect consequently the density of
states is decreased in low energy modes. This removing  the symmetry around
the Fermi level and shifting it, this effect increases with increasing NNN
hopping. If the NNN hopping become competitive with NN hopping the dispersion
will changing complectly as seen in Figure \ref{square2}.
\begin{figure}[h!]
  \centering
  \begin{tabular}{cc}
\includegraphics[scale=.6]{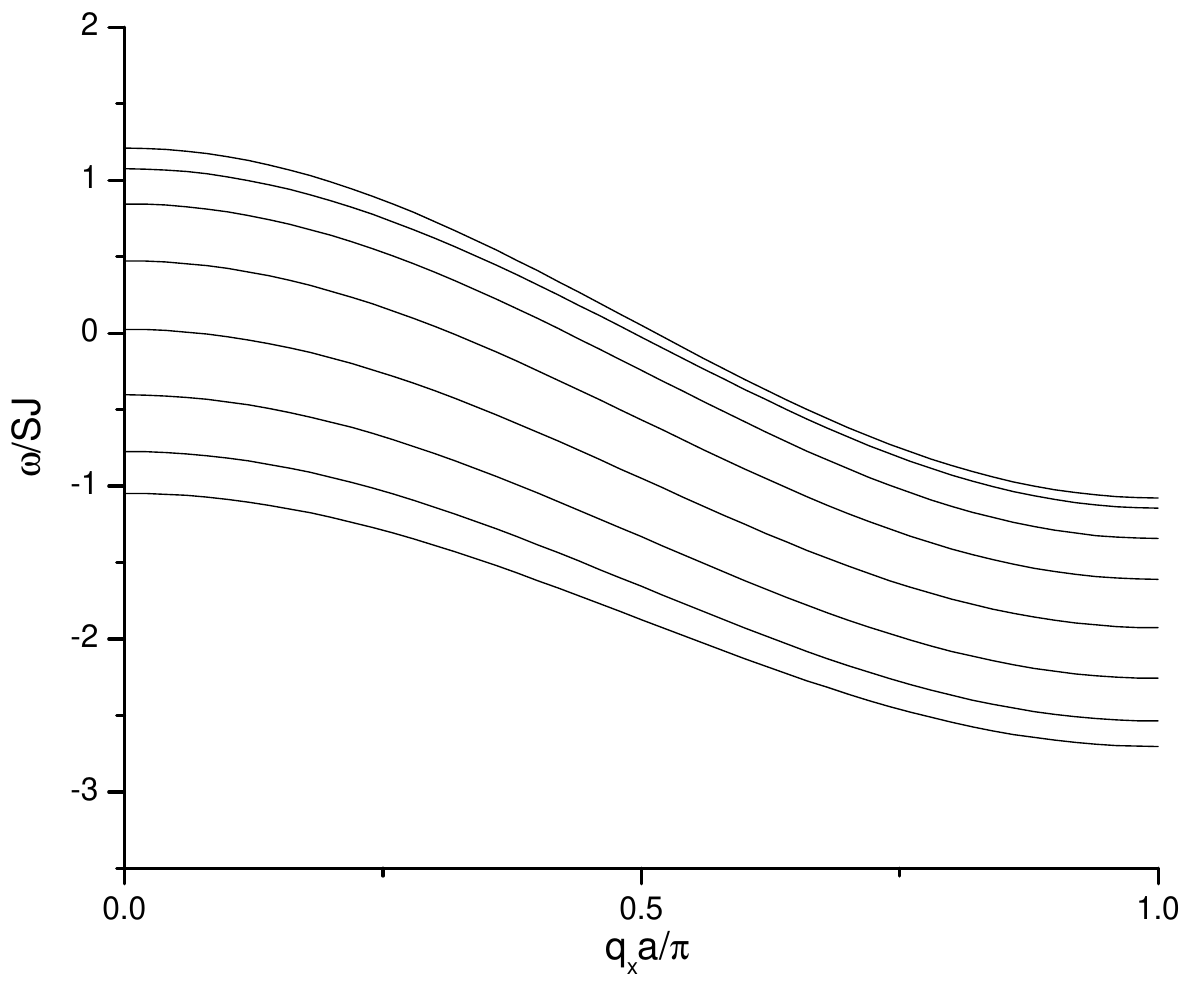}& \includegraphics[scale=.6]{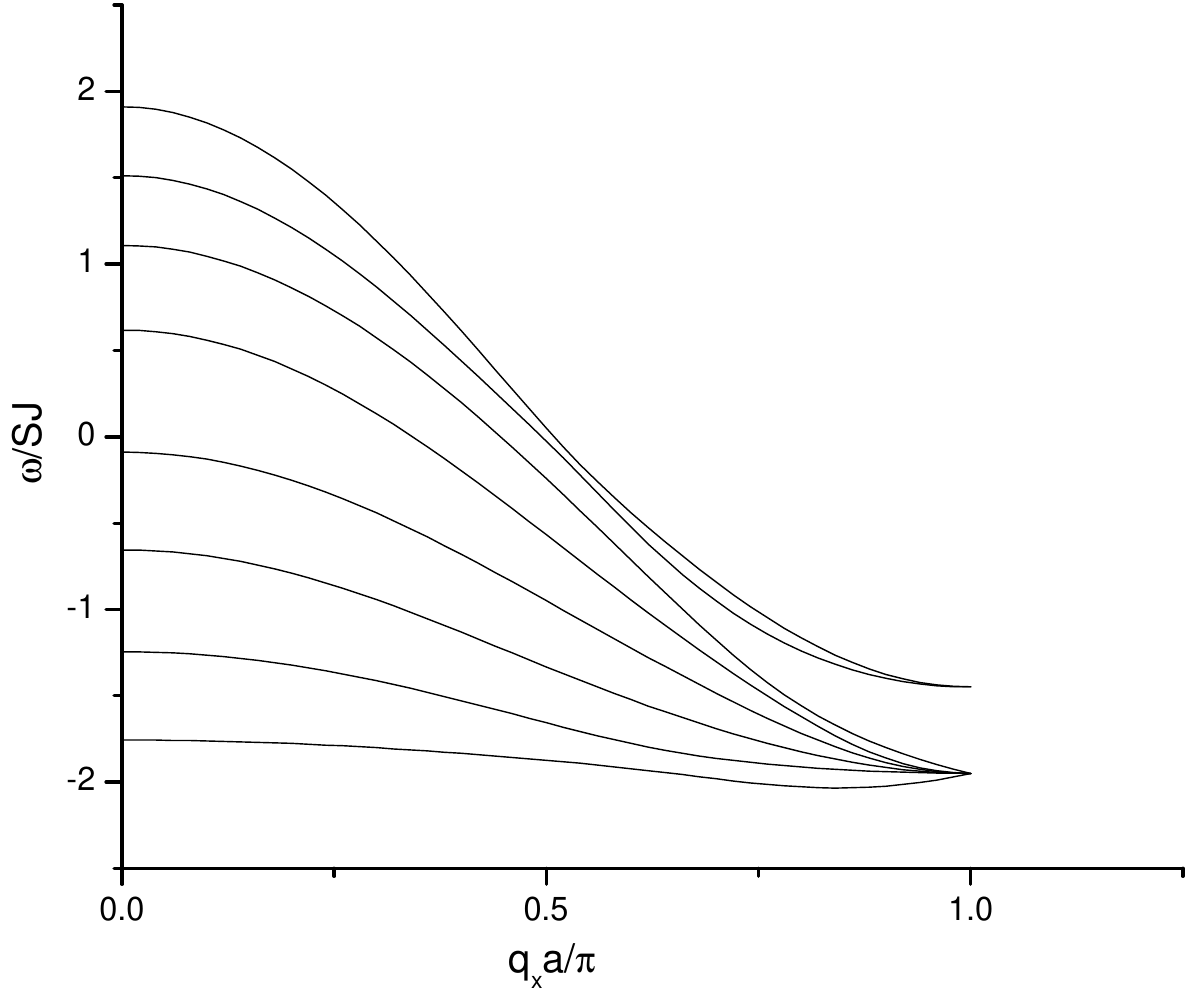}
\end{tabular}
  \caption{The effect of next nearest
neighbor interaction in the dispersion relations on magnetic 2D square lattice with width $N=8$. Right side $t'=0.1t$ left side $t'=0.5t$.}\label{square}
\end{figure}

The above results for NNN hopping is applied to 2D square lattices as shown
in Figure \ref{square} by adding the term $(SJ'/2) (2\cos(q_x a))$ in upper
and lower off diagonal of its $\mathbf{E}$ matrix. The result show the same
behavior for the density of states.

The comparison between the results of $t'=0.1t$ between 2D honeycomb lattice
and 2D square lattice show that the sensitivity for NNN hopping effect is
much larger in the 2D honeycomb lattice than 2D square lattice, this due to
the fact that the number of NNN sites is equal to six which is the double of
NN sites in the 2D honeycomb lattice, while the number of NNN sites is equal
to four which is equal to NN sites in 2D square lattice. Therefore by
changing the ratio between NNN and NN sites in the 2D lattice one can tune
the sensitivity for NNN hopping effects.

\begin{acknowledgments}
This research has been supported by the Egyptian Ministry of Higher Education
and Scientific Research (MZA).
\end{acknowledgments}

\bibliography{xbib2}

\end{document}